\documentclass[aps,pra,showpacs,superscriptaddress,amssymb,twocolumn]{revtex4-1}
\usepackage{graphicx}
\usepackage{appendix} 
\usepackage{bm,amsmath}
\usepackage{dcolumn}

\begin{document}

\vspace{0.2in}

\title{Noisy intermediate-scale quantum computation with 
a complete graph of superconducting qubits: Beyond the single-excitation subspace}

\author{Michael R. Geller}
\email{\tt $\!$mgeller@uga.edu}
\affiliation{Center for Simulational Physics, University of Georgia, Athens, Georgia 30602 USA}
\affiliation{Department of Physics and Astronomy, University of Georgia, Athens, Georgia 30602, USA}

\date{\today}

\begin{abstract}
There is currently a tremendous interest in developing practical applications of noisy intermediate-scale quantum processors without the overhead required by full error correction. Quantum information processing is especially challenging within the gate model, as algorithms quickly lose fidelity as the problem size and circuit depth grow. This has lead to a number of non-gate-model approaches such as analog quantum simulation and quantum annealing. These approaches come with specific hardware requirements that are typically different than that of a universal gate-based quantum computer.
We have previously proposed a non-gate-model approach called the single-excitation subspace (SES) method $[{\rm Phys.~Rev.~A} \ {\bf 91}, 062309 \, (2015)]$, which requires a complete graph of superconducting qubits with tunable coupling. Like any approach lacking error correction, the SES method is not scalable, but it often leads to algorithms with {\it constant} depth, allowing it to outperform the gate model in a wide variety of applications. A challenge of the SES method is that it requires a physical qubit for every basis state in the computer's Hilbert space. This imposes large resource costs for algorithms using registers of ancillary qubits, as each ancilla would double the required graph size. Here we show how to circumvent this doubling by leaving the SES and reintroducing a tensor product structure in the computational subspace. Specifically, we implement the tensor product of an SES register holding ``data" with one or more ancilla qubits, which are able to independently control arbitrary $n\!\times\!n$ unitary operations on the data in a constant number of steps. This enables a hybrid form of quantum computation where fast SES operations are performed on the data, traditional logic gates and measurements are performed on the ancillas, and controlled-unitaries act between. As an application we give an  ancilla-assisted SES implementation of the quantum linear system solver of Harrow, Hassidim, and Lloyd.
\end{abstract}

\maketitle

\section{INTRODUCTION}
\label{introduction section}

The recent Google quantum supremacy experiment \cite{AruteNat19} 
marks the beginning of an exciting era of quantum technology, where nascent quantum devices have nontrivial computational power and pose a challenge to the extended Church-Turing thesis. A key design feature of the Google experiment, which sampled from the outputs of random circuits, is the short circuit depths required, even for 53 qubits. A second design feature is that demonstrating supremacy does not require high circuit fidelity, only a (statistically significant) nonzero one.

However these features are atypical of most known quantum algorithms and applications. While there has been a lot of effort dedicated to finding other short-depth applications, and expectation that some will be found \cite{McCleanNJP16}, achieving supercomputing power with noisy intermediate-scale quantum (NISQ) devices \cite{180100862} (also called {\it prethreshold} devices \cite{170605413}) appears to be extremely challenging. 

The single-excitation subspace (SES) method \cite{PritchettEtalArxiv1008.0701,GellerMartinisEtalPRA15,Katabarwa&GellerPRA15} is a non-gate-model approach that uses a complete graph of $n$ superconducting qubits and performs quantum computations and simulations in the $n$-dimensional SES, where the system Hamiltonian is directly programmed. This eliminates the need to decompose operations into elementary one- and two-qubit gates, allowing larger computations to be performed with the available coherence time. Symmetric $n \! \times \! n$ unitaries can be implemented in a single fast step \cite{GellerMartinisEtalPRA15}, and nonsymmetric unitaries in three \cite{Katabarwa&GellerPRA15}. These building blocks lead to many algorithms with $O(1)$ depth, which is highly desirable for NISQ applications. The method also enables quantum simulation of $n$-dimensional closed systems with constant depth \cite{PritchettEtalArxiv1008.0701,GellerMartinisEtalPRA15}.

\begin{figure}
{\vskip 0.2in}
\includegraphics[width=8cm]{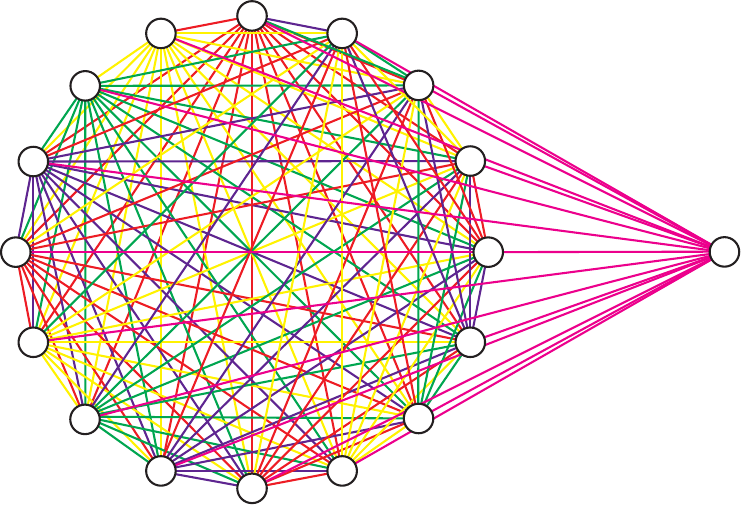} 
\caption{(Color online) Adding a qubit to the $n=16$ graph.}
\label{16plus1 figure}
\end{figure} 

Restriction to the SES means that a physical qubit is required for each basis state of the computational Hilbert space, and the method is not scalable. A technically unscalable architecture, however, might still be useful for practical NISQ computing. But the scaling does impose large resource costs for algorithms involving ancillary qubits, as each ancilla would double the required graph size. Here we show how to circumvent this doubling by reintroducing tensor product structure into the computational Hilbert space, which necessitates leaving the SES. In particular, we show that a complete graph of $n+n'$ qubits can implement the tensor product of an $n$-qubit SES register holding ``data" with $n'$ ancilla qubits, in such a way that each ancilla coherently controls the application of an arbitrary $n \! \times \! n$ unitary to the data. Crucially, the number of steps required to perform a set of $n'$ controlled-unitaries is {\it independent} of $n$ and only linear in $n'$ (as the controlled-unitaries are performed serially). 

To better understand the tensor product structure consider adding a single superconducting qubit to an existing SES array, resulting in a complete graph of $n+1$ qubits; see  Fig.~\ref{16plus1 figure}. There are two distinct ways of doing this, which we call {\it direct sum} and {\it tensor product}. The direct sum means that number of excitations remains unity and the dimension of the computational subspace is increased by one, resulting in an $n+1$-qubit SES register. Adding $n'$ qubits in this way increases the size of the register to $n+n'$ qubits. Or we can say we have added an $n'$-qubit SES register to the original $n$-qubit register. If we denote the computational subspace of an $n$-qubit SES register by ${\rm SES}_{n},$ the direct sum implements
\begin{equation}
{\rm SES}_{n} \oplus {\rm SES}_{n'} = {\rm SES}_{n+n'}.
\label{direct sum identity}
\end{equation}

\begin{figure}
{\vskip 0.2in}
\includegraphics[width=8cm]{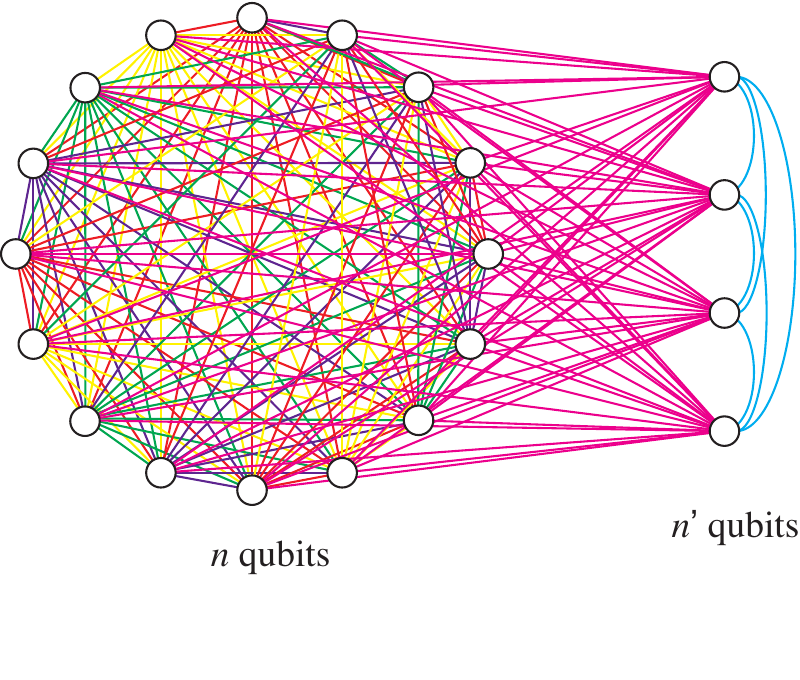} 
\caption{(Color online) Adding a register of $n'$ ancilla qubits creates a hybrid form of SES and gate-based computation, where fast SES operations are performed on the data, traditional logic gates and measurements are performed on the ancillas, and controlled-unitaries operate between.}
\label{16plus4 figure}
\end{figure} 

The tensor product option comes from the standard gate model of quantum computation, where each physical qubit contributes a two-dimensional complex Hilbert space $\mathbb{C}^2$, and the computational Hilbert space of $n$ qubits is the tensor product $\mathbb{C}^2 \otimes \mathbb{C}^2  \otimes \cdots \otimes \mathbb{C}^2.$ In this paper we implement a tensor product of the form
\begin{equation}
{\rm SES}_{n} \otimes \mathbb{C}^2,
\label{tensor product form}
\end{equation}
which adds an excitation and is equivalent in computational subspace size to ${\rm SES}_{2n}.$ The added qubit can be used as an ancilla to control the application of arbitrary unitaries to the data $|\psi\rangle$ stored in ${\rm SES}_{n}$, enabling transformations from product states
\begin{equation}
|\psi\rangle \otimes \big( \alpha |0\rangle_{n+1} + \beta |1\rangle_{n+1} \big)
\label{form of unentangled ancilla}
\end{equation}
to arbitrary states of the form
\begin{equation}
\alpha \, \big( U_0 |\psi\rangle \big) \otimes  |0\rangle_{n+1}
+ \beta \, \big(U_1 |\psi\rangle\big) \otimes  |1\rangle_{n+1}.
\label{form of entangled ancilla}
\end{equation}
Here qubit $n+1$ is the ancilla. Adding $n'$ ancilla in this manner implements
\begin{equation}
{\rm SES}_{n} \otimes \underbrace{ \mathbb{C}^2 \otimes  \cdots \otimes \mathbb{C}^2}_{n' \ {\rm qubits}} = {\rm SES}_{n\times 2^{n'}}.
\label{tensor product identity}
\end{equation}
An example is given in Fig.~\ref{16plus4 figure}. By (\ref{tensor product identity}) we mean that the resulting computational subspace has dimension $n\times 2^{n'} \! .$ Because this is exponential in $n'$, larger problem sizes become possible.  We proposes this as a route to practical quantum computation with NISQ technology.

\section{CONTROLLED-UNITARY PROTOCOL}
\label{controlled unitary section}

\subsection{SES computer chip}

\begin{figure}
{\vskip 0.2in}
\includegraphics[width=6cm]{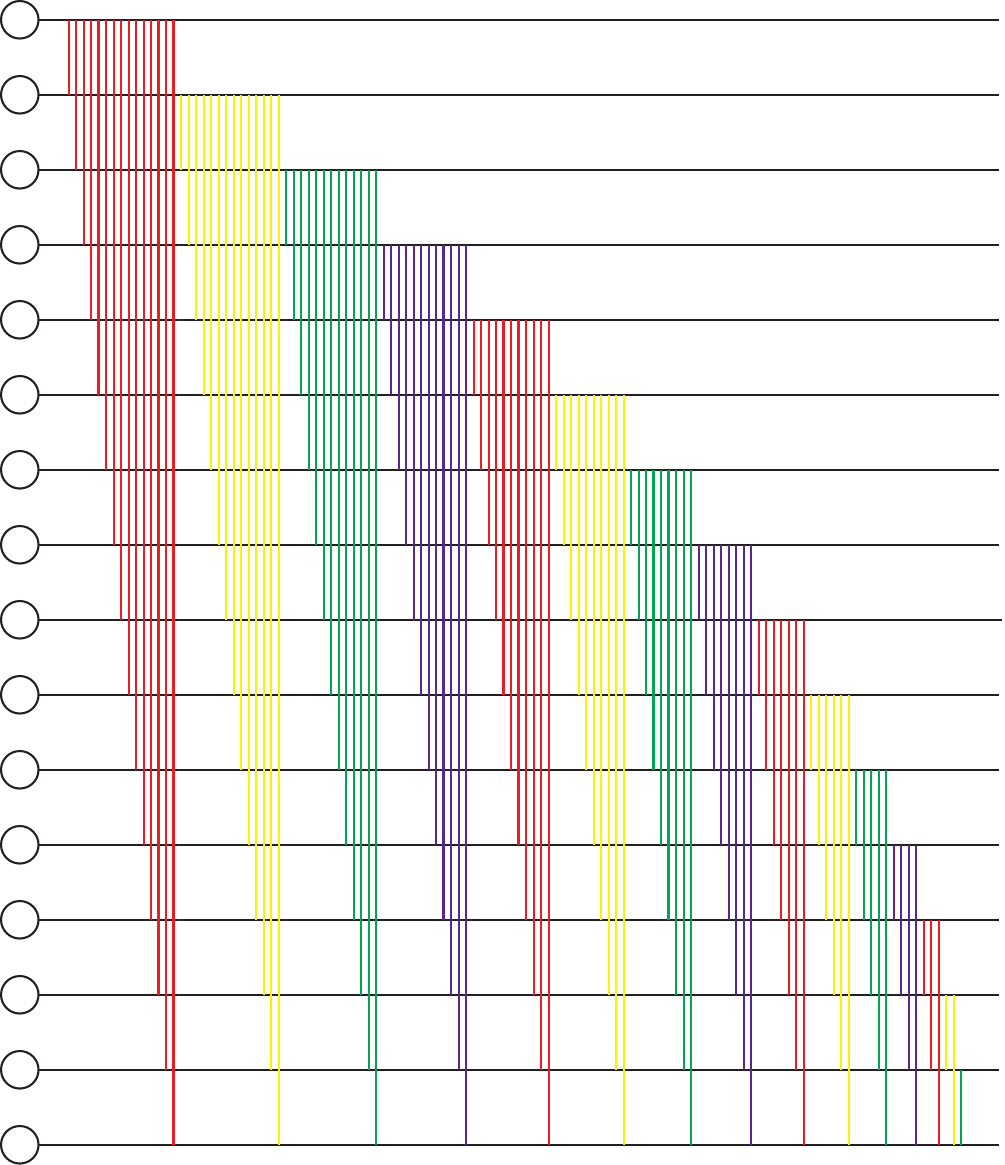} 
\caption{(Color online) Possible layout for an SES chip. The circles represent qubits and the lines are couplers.}
\label{CG16 layout figure}
\end{figure}

The hardware required for ancilla-assisted SES computation is identical to that described in \cite{GellerMartinisEtalPRA15}, i.e., a complete graph (fully connected array) of superconducting transmon \cite{KochPRA07} or Xmon \cite{BarendsPRL13} qubits with tunable frequencies and tunable $\sigma^x \otimes \sigma^x$ couplings \cite{ChenEtalPRL14}. The device Hamiltonian is
\begin{equation}
H_{\rm qc} = \sum_{i} 
\begin{pmatrix}
0 & 0 \\
0 & \epsilon_i  
\end{pmatrix}_{\! \! i}
+ \frac{1}{2} \sum_{i i'} g_{ii'} \, \sigma^x_i\otimes\sigma^x_{i'},
\label{Hqc}
\end{equation}
with $\epsilon_i$ and $g_{ii'}$ tunable. $g_{ii'}$ is a real, symmetric matrix with vanishing diagonal elements. 
A possible chip layout is shown in Fig.~\ref{CG16 layout figure}. 

\subsection{SES method basics}
\label{SES method section}

In the SES method without ancillas \cite{GellerMartinisEtalPRA15,Katabarwa&GellerPRA15}, computations are performed in the $n$-dimensional subspace spanned by the basis states 
\begin{equation}
|i) := |0 \cdots 1_i  \cdots 0\rangle, \ \ \ i \in \lbrace 1, 2, \cdots, n \rbrace.
\label{SES basis states}
\end{equation}
A pure state in the SES has the form 
\begin{equation}
|\psi\rangle = \sum_{i=1}^n a_i \, |i), 
\ \ \ {\rm with} \ \ \ 
\sum_{i=1}^n |a_i|^2 = 1.
\label{SES pure state general form}
\end{equation}
The advantage of working in the SES is that the matrix elements 
\begin{equation}
{\cal H}_{ii'}  := ( i | H_{\rm qc} | i' )  = \epsilon_i \,  \delta_{ii'}  + g_{ii'}
\label{SES hamiltonian}
\end{equation}
of (\ref{Hqc}) can be directly controlled. {\it We can therefore directly program the Hamiltonian of the computer chip.} 

The protocol for implementing a specific operation depends on the functionality (available ranges of the $\epsilon_i$ and $g_{ii'}$) of the SES chip. In this paper we  assume that the experimentally controlled SES Hamiltonian can be written, apart from an additive constant, in the  {\it standard form}
\begin{equation}
{\cal H} = g_{\rm max} K 
\ \ \ {\rm with} \ \  -1 \le K_{i i'} \le 1.
\label{standard form}
\end{equation}
Here $g_{\rm max}$ is the maximum interaction strength provided by the coupler circuits. A reasonable value for $g_{\rm max}/h$ is $10-50 \, {\rm MHz}$.

The basic single-step operation in SES quantum computing is the application of a symmetric unitary of the form $e^{-iA}$ to the data, where $A$ is a given real symmetric matrix. If only the unitary $e^{-iA}$ is known, the classical overhead for obtaining $A$ from $e^{-iA}$ is to be included in the quantum runtime. (Note that the generator $A$ is not unique because the matrix logarithm is not unique.)  Define
\begin{equation}
\theta_{\!A} := \max_{i i'} |A_{ii'} - c \delta_{ii'}|,
\label{angle}
\end{equation}
where $c = (\min_{i} A_{ii}+\max_{i} A_{ii})/2$. The optimal SES Hamiltonian ${\cal H}$ to implement $e^{-iA}$ up to a phase is given by the standard form (\ref{standard form}), with 
\begin{equation}
K = \frac{A-cI}{\theta_{\!A}}.
\label{standard form K}
\end{equation}
Here $I$ is the $n \times n$ identity, and the matrix elements of (\ref{standard form K}) satisfy $|K_{ii'}|\le 1$. The associated evolution time is
\begin{equation}
t_{A}= \frac{\hbar \theta_{\!A}}{g_{\rm max}}. 
\label{standard form evolution time}
\end{equation}
Additional discussion of these results is provided in 
Refs.~\cite{GellerMartinisEtalPRA15} and \cite{Katabarwa&GellerPRA15}.

In an experimental implementation, then, the operation $e^{-iA}$ results from evolution under the Hamiltonian $H_{\rm qc}$ with $\epsilon_i \!=\! \epsilon_0 + g_{\rm max} K_{ii}$, where $\epsilon_0$ is a fixed qubit idling frequency, and $g_{i i'} \!=\! g_{\rm max} K_{ii'} \, (i\neq i'),$ for a time duration $t_A$. It is not even necessary for ${\cal H}$ to be abruptly switched on and off: Any SES Hamiltonian of the form ${\cal H} = g(t) K$ such that $\int (g/\hbar) \, dt = \theta_{\!A}$ may be used.

\subsection{Single-hole states}

The idea underlying the controlled-unitary protocol is to use the non-SES states
\begin{equation}
{\overline{|i)}} := \big(\sigma^x\big)^{\! \otimes n}  |i) = |1 \cdots 1 0_i 1 \cdots 1\rangle,
\label{dual basis states}
\end{equation}
which have $n-1$ excitations and which are particle-hole dual to the SES basis states. The dual state ${\overline{|i)}}$ has a single hole (absence of excitation) in qubit $i$. In a graph with $g_{ii'}=0$, the basis state $|i)$ is an eigenstate with energy $\epsilon_i,$ whereas ${\overline{|i)}}$ has energy $E_n -\epsilon_i$, where
 \begin{equation}
 E_n = \sum_{i=1}^n \epsilon_{i}
\label{En definition}
\end{equation}
is the energy of the filled ``band" $|1 1 \cdots 1\rangle$ of $n$ excitations. Therefore, apart from a shift $E_n$, the dual states have {\it negative} energies; the resulting minus sign is the key to the protocol. 

\subsection{Description of the protocol}
\label{description of the protocol section}

First we discuss the use of a single ancilla. The objective is to implement the controlled-unitary
\begin{equation}
U \otimes |0\rangle \langle 0 |_{n+1} +  I \otimes |1\rangle \langle 1 |_{n+1},
\label{controlled unitary definition}
\end{equation}
where $U$ is an arbitrary $n \! \times \! n$ unitary matrix acting on the SES register, and $I$ is the $n \! \times \! n$ identity. (This definition differs from the usual one by {\sf NOT} gates on the ancilla; we assume that the additional {\sf NOT} gates are included in the complete protocol.) Partition an $n+1$-qubit complete graph into an $n$-qubit SES register and one ancilla.  The initial state is of the form 
\begin{equation}
|\psi\rangle \otimes \big( \alpha \, |0\rangle_{n+1} 
+  \beta \, |1\rangle_{n+1} \big)
\label{abstract initial state}
\end{equation}
or
\begin{equation}
\bigg( \sum_{i=1}^n a_i \, |i) \bigg)
\otimes \big( \alpha |0\rangle_{n+1} + \beta |1\rangle_{n+1} \big).
\label{initial n+1 state}
\end{equation}
Next write the unitary in (\ref{controlled unitary definition}) in spectral form as $V e^{-iD} V^\dagger$, or equivalently
\begin{equation}
U  = V e^{-iD/2} e^{-iD/2} V^\dagger,
\label{special U decomposition}
\end{equation}
where $V$ is unitary and $D$ is a real diagonal matrix. We will make $U$ conditional by implementing
\begin{equation}
U  = V e^{-iD/2} e^{\pm iD/2} V^\dagger
\label{conditional U decomposition}
\end{equation}
instead of (\ref{special U decomposition}), where the plus sign comes from the negative energy of the single-hole states and results in an application of the identity. 

\begin{figure}
{\vskip 0.2in}
\includegraphics[width=4cm]{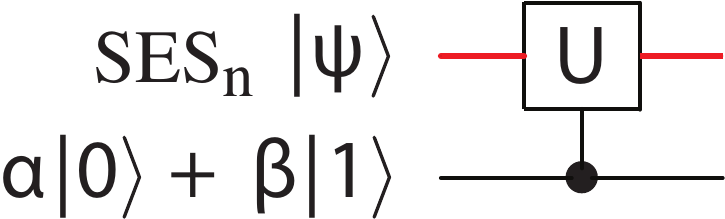} 
\caption{(Color online) Controlled unitary operation. The red upper line is an $n$-qubit SES register, the lower line is the ancilla qubit.}
\label{tensor product circuit figure}
\end{figure} 

The first three steps of the protocol are to implement the $V^\dagger$ operation in (\ref{conditional U decomposition}) on the data stored in the SES register. 
The $KAK$ decomposition \cite{Katabarwa&GellerPRA15} is used to write $V$ and $V^\dagger$ as
\begin{equation}
V = e^{-iA} e^{-iB}  e^{iA} 
\ \ {\rm and} \ \ 
V^\dagger = e^{-iA} e^{iB}  e^{iA},
\end{equation}
where $A$ and $B$ are real symmetric matrices. The procedure for computing $A$ and $B$ is given in \cite{Katabarwa&GellerPRA15}. Each operator produced by the $KAK$ decomposition is a symmetric unitary and can be implemented in a single step (Sec.~\ref{SES method section}). The first operation in the protocol, $e^{iA}$, results from evolution under $H_{\rm qc}$ with $\epsilon_i \!=\! \epsilon_0 + g_{\rm max} K_{ii}$ ($\epsilon_0$ is a fixed qubit idling frequency) and $g_{i \neq i'} \!=\! g_{\rm max} K_{ii'},$ with $K=-(A-cI)/\theta_{\!A},$ for a time duration $t_A\!=\! \hbar \theta_{\!A} /g_{\rm max}$. 
The indices $i$ and $i' $ in these expressions include the SES partition $\lbrace 1,\cdots,n\rbrace$ only, and during this operation all couplings to the ancilla are turned off ($g_{i,n+1}=0$ for all $i \in \lbrace 1,\cdots,n\rbrace).$ These settings program the SES Hamiltonian ${\cal H} = g_{\rm max} K$ into the chip. The ancilla qubit frequency $\epsilon_{n+1}$ is set to $\epsilon_0$.  The protocol implements the symmetric unitary $e^{iA}$ up to a phase, with that phase chosen to minimize the operation time $t_{\rm A}$. $e^{-iA}$ is implemented by changing $K \rightarrow -K$. $e^{\pm iB}$ are implemented by changing $A \rightarrow B$. The total time required to implement $V^\dagger$ is $t_{V} \! = \! 2t_{A} + t_{B}$. After these steps (\ref{initial n+1 state}) becomes
\begin{equation}
\bigg( \sum_{i=1}^n (V^\dagger a)_i  \, |i) \bigg)
\otimes \big( \alpha |0\rangle_{n+1} + e^{-i \epsilon_0 t_{V}/\hbar} \beta |1\rangle_{n+1} \big),
\label{state after V}
\end{equation}
where, for any unitary $W$ acting on the SES,  we write $\sum_{i'} (i|W|i') \, a_{i'}$ as $(Wa)_i$. Note that the ancilla acquired a relative phase after these operations; we assume that such phases are removed by applying $z$ rotations to the ancilla or by working in a rotating frame.

The next steps apply $e^{\pm i D/2}$ conditioned on the ancilla, the sign change resulting from the negative energies of the dual states. After {\sf CNOT} gates between the ancilla (control) and each of the $n$ SES qubits (targets), we have
\begin{equation}
 \sum_{i=1}^n \bigg( \! (V^\dagger a)_i \, |i) \otimes \alpha \,
|0\rangle_{n+1} 
+  (V^\dagger a)_i  \, {\overline{|i)}} \otimes  \beta \, |1\rangle_{n+1} \! \bigg).
\label{state after CNOTs}
\end{equation}
In Sec.~\ref{cnot section} we show to implement these $n$  {\sf CNOT} gates {\it simultaneously}. Then follow the protocol as if to implement the diagonal operator $e^{-iD/2}$ in the SES: Apply $H_{\rm qc}$ with $g_{i i'}\!=\!0$ and $\epsilon_i = \epsilon_0 + g_{\rm max} K_{ii}$ for a time  $t_D\!=\!\hbar \theta_{\!D} / 2 g_{\rm max}$. Here $K \!:=\! (D - cI)/\theta_{\!D},$ $\theta_{\!D} \! := \! \max_{i} |D_{ii}-c|$, and $c = (\min_{i} D_{ii}+\max_{i} D_{ii})/2.$ Set the ancilla frequency to $\epsilon_0.$ Following this operation, and another ancilla $z$ rotation,  (\ref{state after CNOTs}) becomes
\begin{equation}
\sum_{i=1}^n \bigg( (e^{-iD/2} V^\dagger a)_i \, |i) \otimes \alpha \,
|0\rangle_{n+1} 
+  ( e^{iD/2} V^\dagger a)_i  \,  {\overline{|i)}}  \otimes  \beta \, |1\rangle_{n+1} \bigg).
\nonumber
\label{state after D/2}
\end{equation}

After a second set of {\sf CNOT} gates and subsequent application of $e^{-iD/2}$ and $V$ to the SES, and a final ancilla $z$ rotation, we obtain
\begin{equation}
 \sum_{i=1}^n \bigg( (Ua)_i \, |i) \otimes \alpha \,
|0\rangle_{n+1} 
+  a_i  \,  |i)\otimes  \beta \, |1\rangle_{n+1} \bigg),
\label{final state}
\end{equation}
or
\begin{equation}
U |\psi\rangle \otimes \alpha \, |0\rangle_{n+1} 
+  |\psi\rangle   \otimes  \beta \, |1\rangle_{n+1},
\label{abstract final state}
\end{equation}
as required. We represent this operation---including implicit {\sf NOT} gates on the ancilla before and after transformation (\ref{controlled unitary definition})---by the circuit diagram of Fig.~\ref{tensor product circuit figure}.

\begin{figure}
\includegraphics[width=9cm]{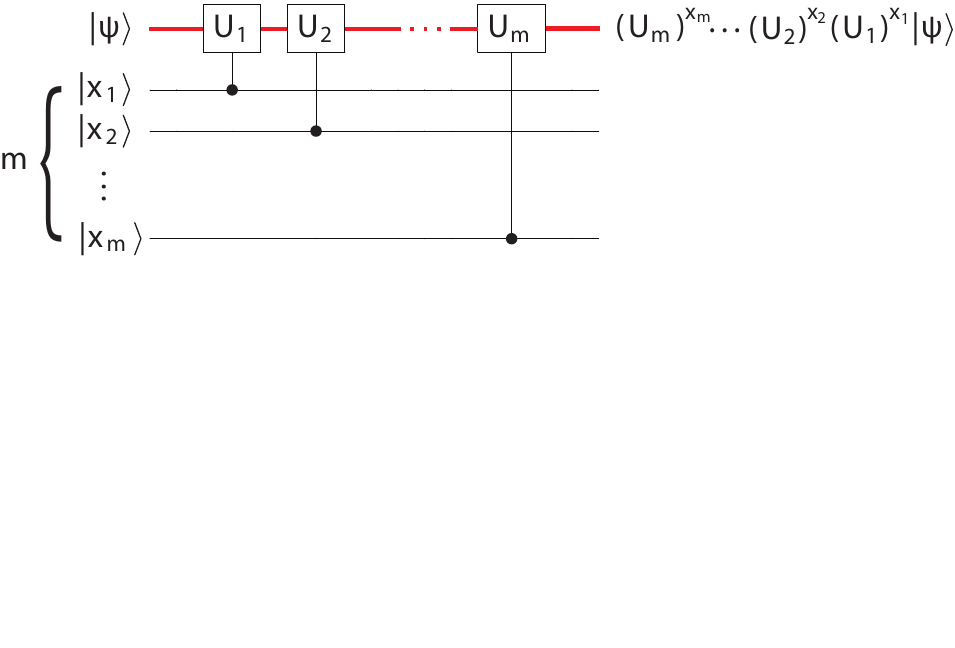} 
\vskip -1.5in
\caption{(Color online) A simple circuit with $m$ distinct controlled unitaries in succession. The red upper line is an $n$-qubit SES register. When the $m$ ancilla are in a computational basis state $| x_1 x_2 \cdots x_m\rangle$ with $x_i \in \lbrace 0, 1\rbrace$, the action on the SES register is $ | \psi \rangle \mapsto  (U_m)^{x_m} \cdots (U_2)^{x_2} (U_1)^{x_1}| \psi \rangle$.}
\label{multiple unitaries circuit figure}
\end{figure} 

We have described the use of a single ancilla qubit. The total number of steps required to implement the controlled unitary (about 10) is independent of $n$. Additional ancilla can be included by increasing the graph size by one for each new ancilla. Each ancilla independently controls unitaries acting on the shared data register. These unitaries, however, cannot be performed simultaneously, so in most applications the runtime will scale linearly with the number of ancilla $n'\! .$  

As an example, in Fig.~\ref{multiple unitaries circuit figure} we consider a quantum circuit that implements $m$ controlled unitaries $U_1, U_2, \dots , U_m$ in succession, each controlled by a single ancilla. This circuit applies the  operator
\begin{equation}
\sum_{x=0}^{{2^m}-1}  (U_m)^{x_m} \cdots (U_2)^{x_2} (U_1)^{x_1} \otimes |x\rangle \langle x |
\label{multiple unitary circuit operator}
\end{equation}
to the computational Hilbert space, where $x_j$ is the $j$th 
bit in the binary representation for $x$, 
\begin{equation}
x = \sum_{j=1}^m 2^{m-j}  x_j.
\end{equation}
A common special case of (\ref{multiple unitary circuit operator}) is to let the $U_j = (U)^{2^{m-j}}$ for some unitary $U$, in which case the circuit of  Fig.~\ref{multiple unitaries circuit figure} implements the operation
\begin{equation}
\sum_{x=0}^{{2^m}-1}  U^x \otimes |x\rangle \langle x |.
\end{equation}

\subsection{Multi-target CNOT}
\label{cnot section}

The protocol of the last section requires two rounds of {\sf CNOT} gates applied between the ancilla (control) and the $n$ qubits in the SES partition (targets).
For small $n$ these can be done serially using the high-fidelity entangling gates developed for standard gate-based superconducting quantum computation. The {\sf CNOT} gates commute, however, and in principal can be performed simultaneously. Multi-target CNOT gate protocols \cite{WangPRL01,YangPRA05b,LinPRA06,YangPRA10,YangPRA10,WaseemPhysicaC12,SongPhysicaB12,YangOL14,LiuPRA14} have been developed for ion trap, cavity QED, and circuit QED architectures, where many qubits can be coupled to a common cavity or other bosonic mode. These protocols can be applied to the complete graph architecture as well, but at the expense of supplementing each ancilla with an additional resonator or qubit, which must also be fully connected. The desired tensor product would then require $n+2n'$ qubits. We avoid this overhead by designing a fast multi-target {\sf CNOT} gate specifically for the complete graph.

It is well known that the entangling gate
\begin{equation}
e^{-i \frac{\pi}{4} \sigma^x \otimes \sigma^x}
\label{XX entangler}
\end{equation}
is {\it locally} equivalent to a two-qubit {\sf CNOT} gate, meaning that it is a {\sf CNOT} apart from single-qubit rotations. To see this, let the second qubit be the control, and apply Hadamards to obtain $e^{-i \frac{\pi}{4} \sigma^x \otimes \sigma^z} \! \! .$ This operator acts with $e^{- i \frac{\pi}{4} \sigma^x}$ on the target when the control is $|0\rangle$, and with $e^{i \frac{\pi}{4} \sigma^x}$ when the control is $|1\rangle$, from which it is straightforward to construct a {\sf CNOT}.

\begin{figure}
{\vskip 0.2in}
\includegraphics[width=8cm]{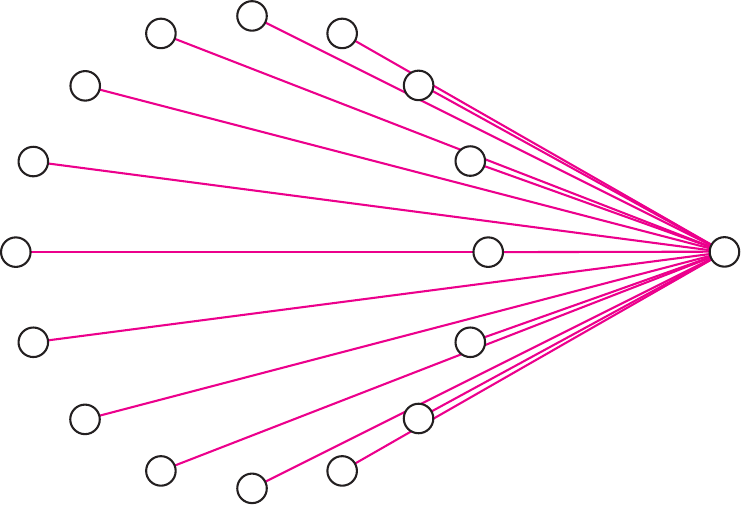} 
\caption{(Color online) Graph to implement the $n$-target {\sf CNOT} gate.}
\label{16plus4forCNOTs figure}
\end{figure} 

It is not surprising that a simultaneous multi-target {\sf CNOT} gate is possible in the complete graph architecture, and the Hamiltonian (\ref{Hqc}) already contains an interaction underlying such an operation: Set the couplings between the ancilla and the $n$ qubits in the SES partition to a positive constant $g$ and all others to zero, as illustrated in Fig.~\ref{16plus4forCNOTs figure}. The interaction 
\begin{equation}
g \left( \sum_{i=1}^n \sigma^x_i \right) \otimes \sigma^x_{n+1}
\label{Hqc interaction}
\end{equation}
couples the ancilla qubit to a collective variable
\begin{equation}
S_x = \sum_{i=1}^n \sigma^x_i 
\label{total spin}
\end{equation}
of the SES partition. Such an interaction, {\it on its own}, can be used to generate the desired multi-qubit entangling operation
\begin{equation}
e^{-i \frac{\pi}{4} S_x \otimes\sigma^x_{n+1}} 
= \prod_{i=1}^n e^{-i \frac{\pi}{4} \sigma^x_i \otimes\sigma^x_{n+1}} 
\label{main entangling gate}
\end{equation}
that generalizes (\ref{XX entangler}). The multi-target {\sf CNOT} gate
\begin{equation}
 I \otimes |0\rangle \langle 0 |_{n+1} +
(\sigma^x)^{\otimes n} \otimes |1\rangle \langle 1|_{n+1} 
\label{multitarget CNOT gate definition}
\end{equation}
results from the gate sequence
\begin{equation}
\begin{pmatrix}
1  & 0 \\
0  & -i \\
\end{pmatrix}_{\! n+1}  \! \! \!
(e^{i \frac{\pi}{4} \sigma^x} )^{\otimes n} 
\, {\sf H}_{n+1} \ e^{-i \frac{\pi}{4} S_x \otimes\sigma^x_{n+1}} \ {\sf H}_{n+1} ,
\label{multitarget CNOT pulse sequence}
\end{equation}
where ${\sf H}$ is the single-qubit Hadamard gate.

The device Hamiltonian (\ref{Hqc}), however, contains single-qubit terms that do not commute with (\ref{Hqc interaction}). Therefore it will be necessary follow a modified protocol to obtain the entangler (\ref{main entangling gate}): Add a $\sigma^x$ microwave drive to the ancilla and transform to the usual rotating frame, where the $\sigma^x \otimes \sigma^x$ interaction becomes $\frac{1}{2}(\sigma^x \otimes \sigma^x + \sigma^y \otimes \sigma^y)$, and then transform to a second rotating frame where the interaction is $\frac{1}{2} \sigma^x \otimes \sigma^x.$ This is discussed further in Appendix \ref{Multi-qubit entangler design section}.

Finally, we discuss the expected performance of this design when implemented in a transmon-based chip with inductive couplers. The main source of error is leakage into higher lying $|2\rangle$ states neglected in (\ref{Hqc}) but present in a real device. Although the current design has not been optimized to minimize this leakage, the estimated performance is already satisfactory for initial demonstrations, as indicated in Table \ref{CNOT gate error table}.

\begin{table}[htb]
\centering
\caption{Performance of simultaneous $n$-target CNOT gate in complete graph of $n\!+\!1$ transmon or Xmon qubits, using realistic models for the qubits and couplers. Here $\eta$ is the qubit anharmonicity, $t_{\rm gate}$ is the gate time excluding the single-qubit rotations in (\ref{multitarget CNOT pulse sequence}), $\Omega$ is the microwave Rabi frequency, and $g$ is the coupler strength. The reported gate error is  $E_{\rm gate} = 1 - |\langle \Psi | U_{\rm ideal} ^\dagger U | \Psi \rangle|^2 \! ,$ where $U_{\rm ideal}$ is the ideal entangler (\ref{main entangling gate}), and $U$ is the realized evolution operator computed in the absence of decoherence. The error is averaged over initial states $|\Psi \rangle$. The qubit frequencies are $\epsilon_0/h = {\rm 5.5 \, GHz}$.}
\begin{tabular}{|c|c|c|c|c|c|}
\hline
$n$ & $\eta/h$ & $t_{\rm gate}$ & $\Omega/h$ & $g/h$ & $E_{\rm gate} $ \\
\hline
\hline
3 & ${\rm 300 \, MHz}$ & ${\rm 40 \, ns}$  & ${\rm 100 \, MHz}$ & ${\rm 6.25 \, MHz}$ & 1.2 \% \\
4 & ${\rm 300 \, MHz}$ & ${\rm 40 \, ns}$  & ${\rm 150 \, MHz}$ & ${\rm 6.25 \, MHz}$ & 1.7 \% \\
5 & ${\rm 300 \, MHz}$ & ${\rm 40 \, ns}$  & ${\rm 150 \, MHz}$ & ${\rm 6.25 \, MHz}$ & 2.4 \% \\
\hline
\end{tabular}
\label{CNOT gate error table}
\end{table}

\section{APPLICATION TO MATRIX INVERSION}

As an application of these techniques we give an ancilla-assisted SES implementation of the quantum linear system solver of Harrow, Hassidim, and Lloyd \cite{HarrowPRL09}. We do not expect an SES chip running this implementation to outperform a classical supercomputer. We choose this algorithm because it requires a large register of ancilla qubits, which is challenging, and because it has interesting generalizations and applications to machine learning.

The matrix inversion algorithm \cite{HarrowPRL09,CladerPRL13} solves the linear system $A {\bf x} = {\bf b}$ for ${\bf x}$, accepting ${\bf b}$ in the form of a normalized pure state $|{\bf b} \rangle$, and returning the solution in the form of a pure state $|{\bf x} \rangle$. In the SES implementation these states are stored in a data register of Hilbert space dimension $n$.  (Note that in our notation $A$ is $n\!\times\!n,$ not $2^n\!\times\!2^n$.) A second register of $m$ qubits is used for the phase estimation subroutine, and one more is used for postselection. The value of $m$ determines the accuracy of the solution. The SES implementation (for symmetric $A$) requires a complete graph of $n+m+1$ qubits. 

\begin{widetext}

\begin{figure}
{\vskip 0.2in}
\includegraphics[width=17cm]{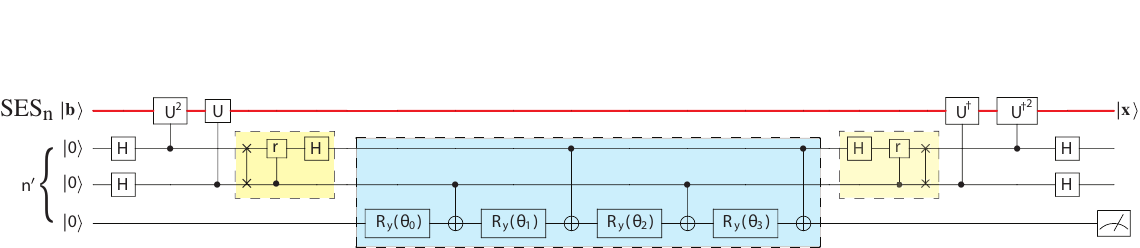} 
\caption{(Color online) Quantum circuit for $n\! \times \! n$ matrix inversion with $m=2$. Here
${\sf H}$ is the Hadamard gate, $U = e^{i A t_0/2^m} \! \! , $ the vertical line connecting crosses is a ${\sf SWAP}$ gate, $r = |0\rangle \langle 0| + i |1\rangle \langle 1|$ is a $z$ rotation, and $R_y = e^{-i(\theta/2) \sigma^y}$ is a $y$ rotation. The small (yellow) subcircuits are the Fourier transforms and the central (blue) subcircuit implements the controlled ancilla rotation (\ref{special contolled rotation}). 
}
\label{matrix inversion circuit figure}
\end{figure} 

\end{widetext}

The circuit for $m=2$ is given in Fig.~\ref{matrix inversion circuit figure}.  The central (blue) subcircuit implements
the controlled-rotation operation
\begin{equation}
\sum_{k=0}^{2^m-1} |k\rangle \langle k | \otimes R_y(\gamma_k), \ \
\gamma_{k} :=
\begin{cases}
2 \arcsin(\frac{1}{k})  & \text{for } k > 0,\\
0 & \text{for } k = 0.
\end{cases}
\label{special contolled rotation}
\end{equation}
Here $|k\rangle$ is a computational basis state of the m-qubit ancilla register. The rotation angles $ \theta_0 ,  \cdots ,  \theta_3 $ in Fig.~\ref{matrix inversion circuit figure} are determined by finding the net $y$ rotation applied to the last qubit in each of the cases $|k\rangle \in \lbrace |00\rangle, |01\rangle, |10\rangle, |11\rangle \rbrace$, making use of the identity $\sigma^x \, R_y(\theta) \, \sigma^x = R_y(-\theta),$ and comparing the result with (\ref{special contolled rotation}), rewritten as
\begin{eqnarray}
&&|00\rangle \langle 00| \otimes R_y(\gamma_0)
+ |01\rangle \langle 01| \otimes R_y(\gamma_1) 
\nonumber \\
&+& |10\rangle \langle 10| \otimes R_y(\gamma_2)
+ |11\rangle \langle 11| \otimes R_y(\gamma_3).
\label{m=2 contolled rotation}
\end{eqnarray}
This leads to 
\begin{equation}
\begin{pmatrix}
1 & 1 & 1 & 1 \\
1 & -1 & -1 & 1 \\
1 & 1 & -1 & -1 \\
1 & -1 & 1 & -1 
\end{pmatrix}
\begin{pmatrix}
\theta_0 \\
\theta_1  \\
\theta_2  \\
\theta_3 
\end{pmatrix}
= 
\begin{pmatrix}
\gamma_0 \\
\gamma_1  \\
\gamma_2  \\
\gamma_3 
\end{pmatrix} \! ,
\label{simultaneous equations for rotation angles}
\end{equation}
with the $\gamma_k$ given in (\ref{special contolled rotation}). The matrix in (\ref{simultaneous equations for rotation angles}), after multiplication by $2^{-m/2}$, is orthogonal and hence immediately inverted, yielding the $\theta_k$.

\begin{figure}
\includegraphics[width=7cm]{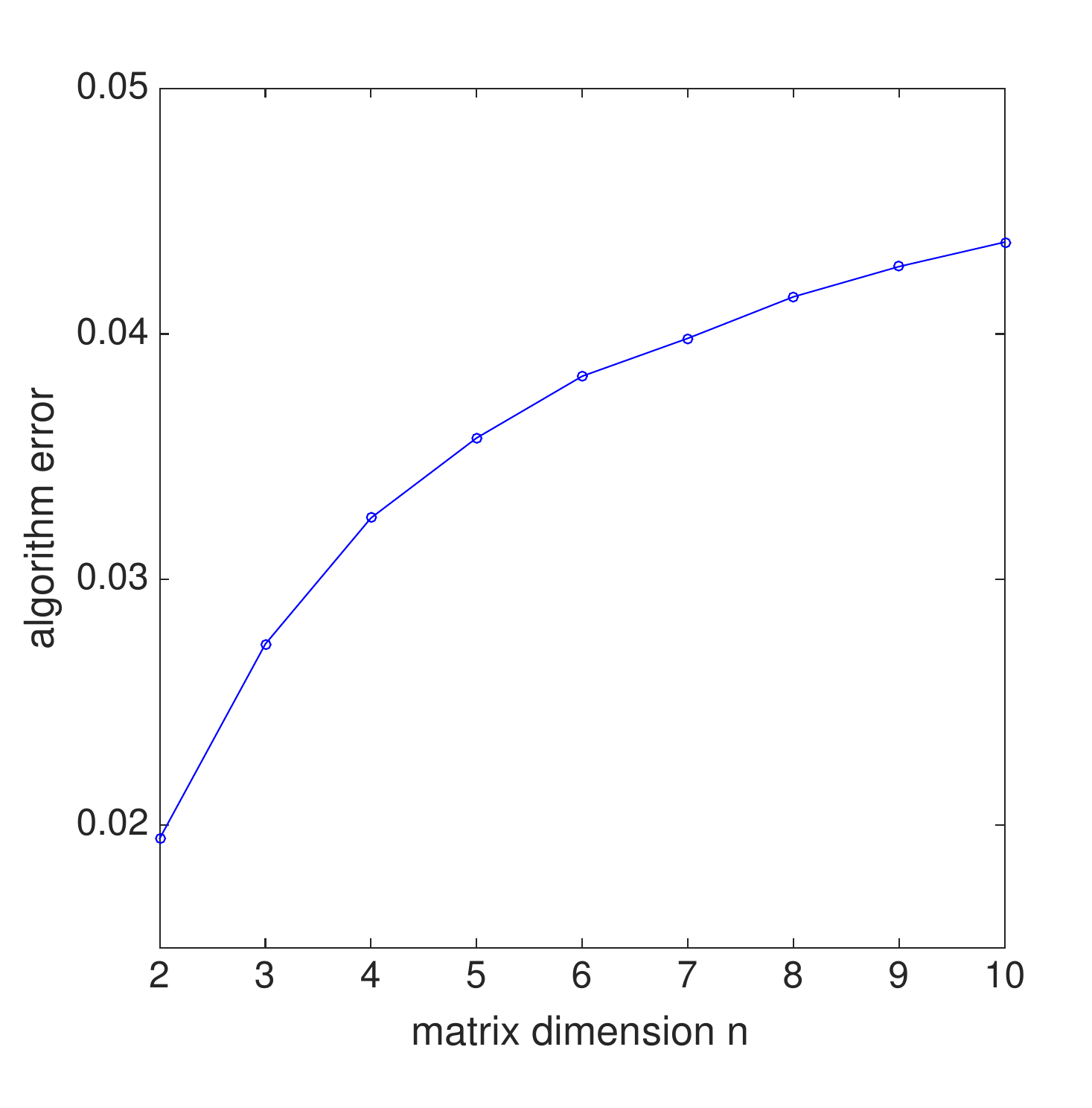} 
\caption{(Color online) Matrix inversion algorithm error for $m=3$, averaged over random real symmetric matrices $A$ (with eigenvalues $0 < \lambda_i < 1$). The error computed here results from the low precision of the phase estimation procedure (small $m$ value) only and does not include the effects of decoherence and other errors that would arise during implementation.}
\label{matrix inversion error figure}
\end{figure}

The $m=2$ phase estimation is not sufficiently accurate for matrix inversion, typically leading to 5-15\% algorithm errors for matrix sizes $2 \le n \le 4.$ By algorithm error we mean 
\begin{equation}
E_{\rm alg} = 1 - \langle {\bf x}_{\rm ideal} | \rho_{\rm data} | {\bf x}_{\rm ideal} \rangle,
\end{equation}
where $ \rho_{\rm data}$ is the final state of the data register, traced over the $m+1$ ancilla, and $| {\bf x}_{\rm ideal} \rangle$ is the pure state corresponding to the exact solution of the given linear system. The circuit of Fig.~\ref{matrix inversion circuit figure} can be easily extended to larger $m$, however, and the performance for $m=3$ is already quite good. The controlled-rotation subcircuit for $m > 2$ can be obtained from the ``uniformly controlled rotation" operator construction of M\"ott\"onen {\it et al.} \cite{MottonenPRL04}, which requires $2^m$ {\sf CNOT} gates (and is therefore not useful for large $m$). Simulating the $m=3$ circuit we find that real symmetric matrices up to dimension 10 can be inverted with algorithm errors less than 5\%, as shown in Fig.~\ref{matrix inversion error figure}, a considerable increase in problem size over the existing gate-based realizations \cite{CaiPRL13,PanPRA14,BarzSciRep14,ZhenPRL17,LeeSciRep19}.

\section{CONCLUSIONS}
\label{conclusion section}

In this paper we have introduced a hybrid form of NISQ computation that extends the reach of the standard SES method \cite{GellerMartinisEtalPRA15,Katabarwa&GellerPRA15} by allowing the use of ancilla as control qubits, without doubling the graph size for each ancilla. This will make it possible to apply the SES method to larger problem sizes. In the applications considered in \cite{PritchettEtalArxiv1008.0701,GellerMartinisEtalPRA15,Katabarwa&GellerPRA15}, as well as the matrix inversion application considered here, the SES method outperforms the standard gate model in terms of the largest problem size that can be implemented with the available coherence time. This makes the SES method useful for practical NISQ applications before the development of fault-tolerant universal quantum computers.
 
\acknowledgements

This work was supported by the National Science Foundation under CDI grant DMR-1029764. It is a pleasure to thank Amara Katabarwa for useful discussions.

\appendix

\section{Multi-qubit entangler design}
\label{Multi-qubit entangler design section}

In this Appendix we show how to produce the entangler (\ref{main entangling gate}) used to construct the multi-target {\sf CNOT} gate. With the couplings between the ancilla and the $n$ qubits in the SES partition set to a positive constant $g$ (see Fig.~\ref{16plus4forCNOTs figure}), the device Hamiltonian  becomes
\begin{equation}
H = \sum_{i=1}^{n+1} 
\begin{pmatrix}
0 & 0 \\
0 & \epsilon_0  
\end{pmatrix}_{\! \! i}
+ g \sum_{i=1}^n \, \sigma^x_i \otimes \sigma^x_{n+1}
+ \Omega \cos( \textstyle{ \frac{\epsilon_0 t}{\hbar} } ) \,  \sigma^x_{n+1},
\label{H with mw}
\end{equation}
where we have added a resonant microwave drive to the ancilla. Decompose the time evolution operator $U$ generated by (\ref{H with mw}) as
\begin{equation}
U = e^{-i D_a t/\hbar} U_a, \ \ {\rm with} \ \ 
D_a = \sum_{i=1}^{n+1} 
\begin{pmatrix}
0 & 0 \\
0 & \epsilon_0  
\end{pmatrix}_{\! \! i} .
\label{a frame}
\end{equation}
Then $\dot{U_a} = -(i/\hbar) H_a U_a,$ where
\begin{equation}
H_a \approx \frac{g}{2}  \sum_{i=1}^n \bigg(  \sigma^x_i  \otimes\sigma^x_{n+1} + \sigma^y_i  \otimes \sigma^y_{n+1} 
\bigg)
+ \frac{\Omega}{2} \sigma^x_{n+1} \ \ \
\label{Ha}
\end{equation}
is the Hamiltonian in the $z$-rotating frame. We choose the evolution time to satisfy
\begin{equation}
t_{\rm gate} = l_a \bigg(\frac{2 \pi \hbar}{\epsilon_0}\bigg), 
\label{gate time quantization condition}
\end{equation}
where $l_a$ is an integer,  which makes $e^{-i D_a t/\hbar} = I$, and which also suppresses the small corrections to (\ref{Ha}) when averaged over $t_{\rm gate}$. The value of $l_a$ (usually  between 100 and 300) is determined by the desired gate time.

Next decompose $U_a$ as
\begin{equation}
U_a = e^{-i D_b t/\hbar} U_b, \ \ {\rm with} \ \ 
D_b = \frac{\Omega}{2} \sigma^x_{n+1}.
\label{b frame}
\end{equation}
Then $\dot{U_b} = -(i/\hbar) H_b U_b,$ where
\begin{equation}
H_b \approx \frac{g}{2}  \sum_{i=1}^n \sigma^x_i  \otimes \sigma^x_{n+1} 
= \frac{g}{2}  S_x \otimes \sigma^x_{n+1} 
\label{Hb}
\end{equation}
is the Hamiltonian in a second frame rotating the ancilla about the $x$ axis.
We choose the Rabi frequency to satisfy
\begin{equation}
\Omega = l_b \bigg(\frac{4 \pi \hbar}{t_{\rm gate}}\bigg), 
\label{gate time quantization condition}
\end{equation}
where $l_b$ is another integer, which makes the corrections to (\ref{Hb}) vanish on average and also makes $e^{-i D_b t/\hbar} = \pm I$. The value of $l_b$ is chosen to minimize the total gate error:  For the simulations reported in Table \ref{CNOT gate error table} we find that $l_b=2$ or 3 is optimal. The effect of this second transformation is to remove the $ \sigma^y \otimes \sigma^y$ term in (\ref{Hb}). Note the factor of $\frac{1}{2}$ in (\ref{Hb}) that is not present in (\ref{H with mw}).

With these parameters the evolution operator becomes
\begin{equation}
U \approx e^{-i \frac{g}{2} S_x \otimes \sigma^x_{n+1}  t_{\rm gate}}.
\label{final approximate evolution operator}
\end{equation}
Setting the coupling strength to $g =\pi \hbar/2 t_{\rm gate}$ generates the desired entangler (\ref{main entangling gate}).

\bibliography{/Users/mgeller/Dropbox/bibliographies/algorithms,/Users/mgeller/Dropbox/bibliographies/applications,/Users/mgeller/Dropbox/bibliographies/dwave,/Users/mgeller/Dropbox/bibliographies/control,/Users/mgeller/Dropbox/bibliographies/error_correction,/Users/mgeller/Dropbox/bibliographies/general,/Users/mgeller/Dropbox/bibliographies/group,/Users/mgeller/Dropbox/bibliographies/ions,/Users/mgeller/Dropbox/bibliographies/math,/Users/mgeller/Dropbox/bibliographies/nmr,/Users/mgeller/Dropbox/bibliographies/optics,/Users/mgeller/Dropbox/bibliographies/simulation,/Users/mgeller/Dropbox/bibliographies/superconductors,/Users/mgeller/Dropbox/bibliographies/surface_code,endnotes}

\begin{thebibliography}{26}%
\makeatletter
\providecommand \@ifxundefined [1]{%
 \@ifx{#1\undefined}
}%
\providecommand \@ifnum [1]{%
 \ifnum #1\expandafter \@firstoftwo
 \else \expandafter \@secondoftwo
 \fi
}%
\providecommand \@ifx [1]{%
 \ifx #1\expandafter \@firstoftwo
 \else \expandafter \@secondoftwo
 \fi
}%
\providecommand \natexlab [1]{#1}%
\providecommand \enquote  [1]{``#1''}%
\providecommand \bibnamefont  [1]{#1}%
\providecommand \bibfnamefont [1]{#1}%
\providecommand \citenamefont [1]{#1}%
\providecommand \href@noop [0]{\@secondoftwo}%
\providecommand \href [0]{\begingroup \@sanitize@url \@href}%
\providecommand \@href[1]{\@@startlink{#1}\@@href}%
\providecommand \@@href[1]{\endgroup#1\@@endlink}%
\providecommand \@sanitize@url [0]{\catcode `\\12\catcode `\$12\catcode
  `\&12\catcode `\#12\catcode `\^12\catcode `\_12\catcode `\%12\relax}%
\providecommand \@@startlink[1]{}%
\providecommand \@@endlink[0]{}%
\providecommand \url  [0]{\begingroup\@sanitize@url \@url }%
\providecommand \@url [1]{\endgroup\@href {#1}{\urlprefix }}%
\providecommand \urlprefix  [0]{URL }%
\providecommand \Eprint [0]{\href }%
\providecommand \doibase [0]{http://dx.doi.org/}%
\providecommand \selectlanguage [0]{\@gobble}%
\providecommand \bibinfo  [0]{\@secondoftwo}%
\providecommand \bibfield  [0]{\@secondoftwo}%
\providecommand \translation [1]{[#1]}%
\providecommand \BibitemOpen [0]{}%
\providecommand \bibitemStop [0]{}%
\providecommand \bibitemNoStop [0]{.\EOS\space}%
\providecommand \EOS [0]{\spacefactor3000\relax}%
\providecommand \BibitemShut  [1]{\csname bibitem#1\endcsname}%
\let\auto@bib@innerbib\@empty
\bibitem [{\citenamefont {Arute}\ \emph {et~al.}(2019)\citenamefont {Arute},
  \citenamefont {Arya}, \citenamefont {Babbush}, \citenamefont {Bacon},
  \citenamefont {Bardin}, \citenamefont {Barends}, \citenamefont {Biswas},
  \citenamefont {Boixo}, \citenamefont {Brandao}, \citenamefont {Buell},
  \citenamefont {Burkett}, \citenamefont {Chen}, \citenamefont {Chen},
  \citenamefont {Chiaro}, \citenamefont {Collins}, \citenamefont {Courtney},
  \citenamefont {Dunsworth}, \citenamefont {Farhi}, \citenamefont {Foxen},
  \citenamefont {Fowler}, \citenamefont {Gidney}, \citenamefont {Giustina},
  \citenamefont {Graff}, \citenamefont {Guerin}, \citenamefont {Habegger},
  \citenamefont {Harrigan}, \citenamefont {Hartmann}, \citenamefont {Ho},
  \citenamefont {Hoffmann}, \citenamefont {Huang}, \citenamefont {Humble},
  \citenamefont {Isakov}, \citenamefont {Jeffrey}, \citenamefont {Jiang},
  \citenamefont {Kafri}, \citenamefont {Kechedzhi}, \citenamefont {Kelly},
  \citenamefont {Klimov}, \citenamefont {Knysh}, \citenamefont {Korotkov},
  \citenamefont {Kostritsa}, \citenamefont {Landhuis}, \citenamefont
  {Lindmark}, \citenamefont {Lucero}, \citenamefont {Lyakh}, \citenamefont
  {Mandr{\`a}}, \citenamefont {McClean}, \citenamefont {McEwen}, \citenamefont
  {Megrant}, \citenamefont {Mi}, \citenamefont {Michielsen}, \citenamefont
  {Mohseni}, \citenamefont {Mutus}, \citenamefont {Naaman}, \citenamefont
  {Neeley}, \citenamefont {Neill}, \citenamefont {Niu}, \citenamefont {Ostby},
  \citenamefont {Petukhov}, \citenamefont {Platt}, \citenamefont {Quintana},
  \citenamefont {Rieffel}, \citenamefont {Roushan}, \citenamefont {Rubin},
  \citenamefont {Sank}, \citenamefont {Satzinger}, \citenamefont {Smelyanskiy},
  \citenamefont {Sung}, \citenamefont {Trevithick}, \citenamefont
  {Vainsencher}, \citenamefont {Villalonga}, \citenamefont {White},
  \citenamefont {Yao}, \citenamefont {Yeh}, \citenamefont {Zalcman},
  \citenamefont {Neven},\ and\ \citenamefont {Martinis}}]{AruteNat19}%
  \BibitemOpen
  \bibfield  {author} {\bibinfo {author} {\bibfnamefont {F.}~\bibnamefont
  {Arute}}, \bibinfo {author} {\bibfnamefont {K.}~\bibnamefont {Arya}},
  \bibinfo {author} {\bibfnamefont {R.}~\bibnamefont {Babbush}}, \bibinfo
  {author} {\bibfnamefont {D.}~\bibnamefont {Bacon}}, \bibinfo {author}
  {\bibfnamefont {J.~C.}\ \bibnamefont {Bardin}}, \bibinfo {author}
  {\bibfnamefont {R.}~\bibnamefont {Barends}}, \bibinfo {author} {\bibfnamefont
  {R.}~\bibnamefont {Biswas}}, \bibinfo {author} {\bibfnamefont
  {S.}~\bibnamefont {Boixo}}, \bibinfo {author} {\bibfnamefont {F.~G. S.~L.}\
  \bibnamefont {Brandao}}, \bibinfo {author} {\bibfnamefont {D.~A.}\
  \bibnamefont {Buell}}, \bibinfo {author} {\bibfnamefont {B.}~\bibnamefont
  {Burkett}}, \bibinfo {author} {\bibfnamefont {Y.}~\bibnamefont {Chen}},
  \bibinfo {author} {\bibfnamefont {Z.}~\bibnamefont {Chen}}, \bibinfo {author}
  {\bibfnamefont {B.}~\bibnamefont {Chiaro}}, \bibinfo {author} {\bibfnamefont
  {R.}~\bibnamefont {Collins}}, \bibinfo {author} {\bibfnamefont
  {W.}~\bibnamefont {Courtney}}, \bibinfo {author} {\bibfnamefont
  {A.}~\bibnamefont {Dunsworth}}, \bibinfo {author} {\bibfnamefont
  {E.}~\bibnamefont {Farhi}}, \bibinfo {author} {\bibfnamefont
  {B.}~\bibnamefont {Foxen}}, \bibinfo {author} {\bibfnamefont
  {A.}~\bibnamefont {Fowler}}, \bibinfo {author} {\bibfnamefont
  {C.}~\bibnamefont {Gidney}}, \bibinfo {author} {\bibfnamefont
  {M.}~\bibnamefont {Giustina}}, \bibinfo {author} {\bibfnamefont
  {R.}~\bibnamefont {Graff}}, \bibinfo {author} {\bibfnamefont
  {K.}~\bibnamefont {Guerin}}, \bibinfo {author} {\bibfnamefont
  {S.}~\bibnamefont {Habegger}}, \bibinfo {author} {\bibfnamefont {M.~P.}\
  \bibnamefont {Harrigan}}, \bibinfo {author} {\bibfnamefont {M.~J.}\
  \bibnamefont {Hartmann}}, \bibinfo {author} {\bibfnamefont {A.}~\bibnamefont
  {Ho}}, \bibinfo {author} {\bibfnamefont {M.}~\bibnamefont {Hoffmann}},
  \bibinfo {author} {\bibfnamefont {T.}~\bibnamefont {Huang}}, \bibinfo
  {author} {\bibfnamefont {T.~S.}\ \bibnamefont {Humble}}, \bibinfo {author}
  {\bibfnamefont {S.~V.}\ \bibnamefont {Isakov}}, \bibinfo {author}
  {\bibfnamefont {E.}~\bibnamefont {Jeffrey}}, \bibinfo {author} {\bibfnamefont
  {Z.}~\bibnamefont {Jiang}}, \bibinfo {author} {\bibfnamefont
  {D.}~\bibnamefont {Kafri}}, \bibinfo {author} {\bibfnamefont
  {K.}~\bibnamefont {Kechedzhi}}, \bibinfo {author} {\bibfnamefont
  {J.}~\bibnamefont {Kelly}}, \bibinfo {author} {\bibfnamefont {P.~V.}\
  \bibnamefont {Klimov}}, \bibinfo {author} {\bibfnamefont {S.}~\bibnamefont
  {Knysh}}, \bibinfo {author} {\bibfnamefont {A.}~\bibnamefont {Korotkov}},
  \bibinfo {author} {\bibfnamefont {F.}~\bibnamefont {Kostritsa}}, \bibinfo
  {author} {\bibfnamefont {D.}~\bibnamefont {Landhuis}}, \bibinfo {author}
  {\bibfnamefont {M.}~\bibnamefont {Lindmark}}, \bibinfo {author}
  {\bibfnamefont {E.}~\bibnamefont {Lucero}}, \bibinfo {author} {\bibfnamefont
  {D.}~\bibnamefont {Lyakh}}, \bibinfo {author} {\bibfnamefont
  {S.}~\bibnamefont {Mandr{\`a}}}, \bibinfo {author} {\bibfnamefont {J.~R.}\
  \bibnamefont {McClean}}, \bibinfo {author} {\bibfnamefont {M.}~\bibnamefont
  {McEwen}}, \bibinfo {author} {\bibfnamefont {A.}~\bibnamefont {Megrant}},
  \bibinfo {author} {\bibfnamefont {X.}~\bibnamefont {Mi}}, \bibinfo {author}
  {\bibfnamefont {K.}~\bibnamefont {Michielsen}}, \bibinfo {author}
  {\bibfnamefont {M.}~\bibnamefont {Mohseni}}, \bibinfo {author} {\bibfnamefont
  {J.}~\bibnamefont {Mutus}}, \bibinfo {author} {\bibfnamefont
  {O.}~\bibnamefont {Naaman}}, \bibinfo {author} {\bibfnamefont
  {M.}~\bibnamefont {Neeley}}, \bibinfo {author} {\bibfnamefont
  {C.}~\bibnamefont {Neill}}, \bibinfo {author} {\bibfnamefont {M.~Y.}\
  \bibnamefont {Niu}}, \bibinfo {author} {\bibfnamefont {E.}~\bibnamefont
  {Ostby}}, \bibinfo {author} {\bibfnamefont {A.}~\bibnamefont {Petukhov}},
  \bibinfo {author} {\bibfnamefont {J.~C.}\ \bibnamefont {Platt}}, \bibinfo
  {author} {\bibfnamefont {C.}~\bibnamefont {Quintana}}, \bibinfo {author}
  {\bibfnamefont {E.~G.}\ \bibnamefont {Rieffel}}, \bibinfo {author}
  {\bibfnamefont {P.}~\bibnamefont {Roushan}}, \bibinfo {author} {\bibfnamefont
  {N.~C.}\ \bibnamefont {Rubin}}, \bibinfo {author} {\bibfnamefont
  {D.}~\bibnamefont {Sank}}, \bibinfo {author} {\bibfnamefont {K.~J.}\
  \bibnamefont {Satzinger}}, \bibinfo {author} {\bibfnamefont {V.}~\bibnamefont
  {Smelyanskiy}}, \bibinfo {author} {\bibfnamefont {K.~J.}\ \bibnamefont
  {Sung}}, \bibinfo {author} {\bibfnamefont {M.~D.}\ \bibnamefont
  {Trevithick}}, \bibinfo {author} {\bibfnamefont {A.}~\bibnamefont
  {Vainsencher}}, \bibinfo {author} {\bibfnamefont {B.}~\bibnamefont
  {Villalonga}}, \bibinfo {author} {\bibfnamefont {T.}~\bibnamefont {White}},
  \bibinfo {author} {\bibfnamefont {Z.~J.}\ \bibnamefont {Yao}}, \bibinfo
  {author} {\bibfnamefont {P.}~\bibnamefont {Yeh}}, \bibinfo {author}
  {\bibfnamefont {A.}~\bibnamefont {Zalcman}}, \bibinfo {author} {\bibfnamefont
  {H.}~\bibnamefont {Neven}}, \ and\ \bibinfo {author} {\bibfnamefont {J.~M.}\
  \bibnamefont {Martinis}},\ }\href@noop {} {\bibfield  {journal} {\bibinfo
  {journal} {Nature}\ }\textbf {\bibinfo {volume} {574}},\ \bibinfo {pages}
  {505} (\bibinfo {year} {2019})}\BibitemShut {NoStop}%
\bibitem [{\citenamefont {McClean}\ \emph {et~al.}(2016)\citenamefont
  {McClean}, \citenamefont {Romero}, \citenamefont {Babbush},\ and\
  \citenamefont {Aspuru-Guzik}}]{McCleanNJP16}%
  \BibitemOpen
  \bibfield  {author} {\bibinfo {author} {\bibfnamefont {J.~R.}\ \bibnamefont
  {McClean}}, \bibinfo {author} {\bibfnamefont {J.}~\bibnamefont {Romero}},
  \bibinfo {author} {\bibfnamefont {R.}~\bibnamefont {Babbush}}, \ and\
  \bibinfo {author} {\bibfnamefont {A.}~\bibnamefont {Aspuru-Guzik}},\
  }\href@noop {} {\bibfield  {journal} {\bibinfo  {journal} {New J. Phys.}\
  }\textbf {\bibinfo {volume} {18}},\ \bibinfo {pages} {023023} (\bibinfo
  {year} {2016})}\BibitemShut {NoStop}%
\bibitem [{\citenamefont {Preskill}(2018)}]{180100862}%
  \BibitemOpen
  \bibfield  {author} {\bibinfo {author} {\bibfnamefont {J.}~\bibnamefont
  {Preskill}},\ }\href@noop {} {\bibfield  {journal} {\bibinfo  {journal}
  {Quantum}\ }\textbf {\bibinfo {volume} {2}},\ \bibinfo {pages} {79} (\bibinfo
  {year} {2018})}\BibitemShut {NoStop}%
\bibitem [{\citenamefont {Aspuru-Guzik}\ \emph {et~al.}()\citenamefont
  {Aspuru-Guzik}, \citenamefont {Wasielewski}, \citenamefont {Olson},
  \citenamefont {Cao}, \citenamefont {Romero}, \citenamefont {Johnson},
  \citenamefont {Dallaire-Demers}, \citenamefont {Sawaya}, \citenamefont
  {Narang},\ and\ \citenamefont {Kivlichan}}]{170605413}%
  \BibitemOpen
  \bibfield  {author} {\bibinfo {author} {\bibfnamefont {A.}~\bibnamefont
  {Aspuru-Guzik}}, \bibinfo {author} {\bibfnamefont {M.}~\bibnamefont
  {Wasielewski}}, \bibinfo {author} {\bibfnamefont {J.}~\bibnamefont {Olson}},
  \bibinfo {author} {\bibfnamefont {Y.}~\bibnamefont {Cao}}, \bibinfo {author}
  {\bibfnamefont {J.}~\bibnamefont {Romero}}, \bibinfo {author} {\bibfnamefont
  {P.}~\bibnamefont {Johnson}}, \bibinfo {author} {\bibfnamefont {P.-L.}\
  \bibnamefont {Dallaire-Demers}}, \bibinfo {author} {\bibfnamefont
  {N.}~\bibnamefont {Sawaya}}, \bibinfo {author} {\bibfnamefont
  {P.}~\bibnamefont {Narang}}, \ and\ \bibinfo {author} {\bibfnamefont
  {I.}~\bibnamefont {Kivlichan}},\ }\href@noop {} {\enquote {\bibinfo {title}
  {Quantum information and computation for chemistry: {NSF} workshop report},}\
  }\bibinfo {note} {{arXiv}: 1706.05413}\BibitemShut {NoStop}%
\bibitem [{\citenamefont {Pritchett}\ \emph {et~al.}()\citenamefont
  {Pritchett}, \citenamefont {Benjamin}, \citenamefont {Galiautdinov},
  \citenamefont {Geller}, \citenamefont {Sornborger}, \citenamefont {Stancil},\
  and\ \citenamefont {Martinis}}]{PritchettEtalArxiv1008.0701}%
  \BibitemOpen
  \bibfield  {author} {\bibinfo {author} {\bibfnamefont {E.~J.}\ \bibnamefont
  {Pritchett}}, \bibinfo {author} {\bibfnamefont {C.}~\bibnamefont {Benjamin}},
  \bibinfo {author} {\bibfnamefont {A.}~\bibnamefont {Galiautdinov}}, \bibinfo
  {author} {\bibfnamefont {M.~R.}\ \bibnamefont {Geller}}, \bibinfo {author}
  {\bibfnamefont {A.~T.}\ \bibnamefont {Sornborger}}, \bibinfo {author}
  {\bibfnamefont {P.~C.}\ \bibnamefont {Stancil}}, \ and\ \bibinfo {author}
  {\bibfnamefont {J.~M.}\ \bibnamefont {Martinis}},\ }\href@noop {} {\enquote
  {\bibinfo {title} {Quantum simulation of molecular collisions with
  superconducting qubits},}\ }\bibinfo {note} {{a}rXiv:1008.0701}\BibitemShut
  {NoStop}%
\bibitem [{\citenamefont {Geller}\ \emph {et~al.}(2015)\citenamefont {Geller},
  \citenamefont {Martinis}, \citenamefont {Sornborger}, \citenamefont
  {Stancil}, \citenamefont {Pritchett}, \citenamefont {You},\ and\
  \citenamefont {Galiautdinov}}]{GellerMartinisEtalPRA15}%
  \BibitemOpen
  \bibfield  {author} {\bibinfo {author} {\bibfnamefont {M.~R.}\ \bibnamefont
  {Geller}}, \bibinfo {author} {\bibfnamefont {J.~M.}\ \bibnamefont
  {Martinis}}, \bibinfo {author} {\bibfnamefont {A.~T.}\ \bibnamefont
  {Sornborger}}, \bibinfo {author} {\bibfnamefont {P.~C.}\ \bibnamefont
  {Stancil}}, \bibinfo {author} {\bibfnamefont {E.~J.}\ \bibnamefont
  {Pritchett}}, \bibinfo {author} {\bibfnamefont {H.}~\bibnamefont {You}}, \
  and\ \bibinfo {author} {\bibfnamefont {A.}~\bibnamefont {Galiautdinov}},\
  }\href@noop {} {\bibfield  {journal} {\bibinfo  {journal} {Phys. Rev. A}\
  }\textbf {\bibinfo {volume} {91}},\ \bibinfo {pages} {062309} (\bibinfo
  {year} {2015})}\BibitemShut {NoStop}%
\bibitem [{\citenamefont {Katabarwa}\ and\ \citenamefont
  {Geller}(2015)}]{Katabarwa&GellerPRA15}%
  \BibitemOpen
  \bibfield  {author} {\bibinfo {author} {\bibfnamefont {A.}~\bibnamefont
  {Katabarwa}}\ and\ \bibinfo {author} {\bibfnamefont {M.~R.}\ \bibnamefont
  {Geller}},\ }\href@noop {} {\bibfield  {journal} {\bibinfo  {journal} {Phys.
  Rev. A}\ }\textbf {\bibinfo {volume} {92}},\ \bibinfo {pages} {032306}
  (\bibinfo {year} {2015})}\BibitemShut {NoStop}%
\bibitem [{\citenamefont {Koch}\ \emph {et~al.}(2007)\citenamefont {Koch},
  \citenamefont {Yu}, \citenamefont {Gambetta}, \citenamefont {Houck},
  \citenamefont {Schuster}, \citenamefont {Majer}, \citenamefont {Blais},
  \citenamefont {Devoret}, \citenamefont {Girvin},\ and\ \citenamefont
  {Schoelkopf}}]{KochPRA07}%
  \BibitemOpen
  \bibfield  {author} {\bibinfo {author} {\bibfnamefont {J.}~\bibnamefont
  {Koch}}, \bibinfo {author} {\bibfnamefont {T.~M.}\ \bibnamefont {Yu}},
  \bibinfo {author} {\bibfnamefont {J.}~\bibnamefont {Gambetta}}, \bibinfo
  {author} {\bibfnamefont {A.~A.}\ \bibnamefont {Houck}}, \bibinfo {author}
  {\bibfnamefont {D.~I.}\ \bibnamefont {Schuster}}, \bibinfo {author}
  {\bibfnamefont {J.}~\bibnamefont {Majer}}, \bibinfo {author} {\bibfnamefont
  {A.}~\bibnamefont {Blais}}, \bibinfo {author} {\bibfnamefont {M.~H.}\
  \bibnamefont {Devoret}}, \bibinfo {author} {\bibfnamefont {S.~M.}\
  \bibnamefont {Girvin}}, \ and\ \bibinfo {author} {\bibfnamefont {R.~J.}\
  \bibnamefont {Schoelkopf}},\ }\href@noop {} {\bibfield  {journal} {\bibinfo
  {journal} {Phys. Rev. A}\ }\textbf {\bibinfo {volume} {76}},\ \bibinfo
  {pages} {042319} (\bibinfo {year} {2007})}\BibitemShut {NoStop}%
\bibitem [{\citenamefont {Barends}\ \emph {et~al.}(2013)\citenamefont
  {Barends}, \citenamefont {Kelly}, \citenamefont {Megrant}, \citenamefont
  {Sank}, \citenamefont {Jeffrey}, \citenamefont {Chen}, \citenamefont {Yin},
  \citenamefont {Chiaro}, \citenamefont {Mutus}, \citenamefont {Neill},
  \citenamefont {O$^\prime$~Malley}, \citenamefont {Roushan}, \citenamefont
  {Wenner}, \citenamefont {White}, \citenamefont {Cleland},\ and\ \citenamefont
  {Martinis}}]{BarendsPRL13}%
  \BibitemOpen
  \bibfield  {author} {\bibinfo {author} {\bibfnamefont {R.}~\bibnamefont
  {Barends}}, \bibinfo {author} {\bibfnamefont {J.}~\bibnamefont {Kelly}},
  \bibinfo {author} {\bibfnamefont {A.}~\bibnamefont {Megrant}}, \bibinfo
  {author} {\bibfnamefont {D.}~\bibnamefont {Sank}}, \bibinfo {author}
  {\bibfnamefont {E.}~\bibnamefont {Jeffrey}}, \bibinfo {author} {\bibfnamefont
  {Y.}~\bibnamefont {Chen}}, \bibinfo {author} {\bibfnamefont {Y.}~\bibnamefont
  {Yin}}, \bibinfo {author} {\bibfnamefont {B.}~\bibnamefont {Chiaro}},
  \bibinfo {author} {\bibfnamefont {J.}~\bibnamefont {Mutus}}, \bibinfo
  {author} {\bibfnamefont {C.}~\bibnamefont {Neill}}, \bibinfo {author}
  {\bibfnamefont {P.}~\bibnamefont {O$^\prime$~Malley}}, \bibinfo {author}
  {\bibfnamefont {P.}~\bibnamefont {Roushan}}, \bibinfo {author} {\bibfnamefont
  {J.}~\bibnamefont {Wenner}}, \bibinfo {author} {\bibfnamefont {T.~C.}\
  \bibnamefont {White}}, \bibinfo {author} {\bibfnamefont {A.~N.}\ \bibnamefont
  {Cleland}}, \ and\ \bibinfo {author} {\bibfnamefont {J.~M.}\ \bibnamefont
  {Martinis}},\ }\href@noop {} {\bibfield  {journal} {\bibinfo  {journal}
  {Phys. Rev. Lett.}\ }\textbf {\bibinfo {volume} {111}},\ \bibinfo {pages}
  {080502} (\bibinfo {year} {2013})}\BibitemShut {NoStop}%
\bibitem [{\citenamefont {Chen}\ \emph {et~al.}(2014)\citenamefont {Chen},
  \citenamefont {Neill}, \citenamefont {Roushan}, \citenamefont {Leung},
  \citenamefont {Fang}, \citenamefont {Barends}, \citenamefont {Kelly},
  \citenamefont {Campbell}, \citenamefont {Chen}, \citenamefont {Chiaro},
  \citenamefont {Dunsworth}, \citenamefont {Jeffrey}, \citenamefont {Megrant},
  \citenamefont {Mutus}, \citenamefont {O$^\prime$Malley}, \citenamefont
  {Quintana}, \citenamefont {Sank}, \citenamefont {Vainsencher}, \citenamefont
  {Wenner}, \citenamefont {White}, \citenamefont {Geller}, \citenamefont
  {Cleland},\ and\ \citenamefont {Martinis}}]{ChenEtalPRL14}%
  \BibitemOpen
  \bibfield  {author} {\bibinfo {author} {\bibfnamefont {Y.}~\bibnamefont
  {Chen}}, \bibinfo {author} {\bibfnamefont {C.}~\bibnamefont {Neill}},
  \bibinfo {author} {\bibfnamefont {P.}~\bibnamefont {Roushan}}, \bibinfo
  {author} {\bibfnamefont {N.}~\bibnamefont {Leung}}, \bibinfo {author}
  {\bibfnamefont {M.}~\bibnamefont {Fang}}, \bibinfo {author} {\bibfnamefont
  {R.}~\bibnamefont {Barends}}, \bibinfo {author} {\bibfnamefont
  {J.}~\bibnamefont {Kelly}}, \bibinfo {author} {\bibfnamefont
  {B.}~\bibnamefont {Campbell}}, \bibinfo {author} {\bibfnamefont
  {Z.}~\bibnamefont {Chen}}, \bibinfo {author} {\bibfnamefont {B.}~\bibnamefont
  {Chiaro}}, \bibinfo {author} {\bibfnamefont {A.}~\bibnamefont {Dunsworth}},
  \bibinfo {author} {\bibfnamefont {E.}~\bibnamefont {Jeffrey}}, \bibinfo
  {author} {\bibfnamefont {A.}~\bibnamefont {Megrant}}, \bibinfo {author}
  {\bibfnamefont {J.~Y.}\ \bibnamefont {Mutus}}, \bibinfo {author}
  {\bibfnamefont {P.~J.~J.}\ \bibnamefont {O$^\prime$Malley}}, \bibinfo
  {author} {\bibfnamefont {C.~M.}\ \bibnamefont {Quintana}}, \bibinfo {author}
  {\bibfnamefont {D.}~\bibnamefont {Sank}}, \bibinfo {author} {\bibfnamefont
  {A.}~\bibnamefont {Vainsencher}}, \bibinfo {author} {\bibfnamefont
  {J.}~\bibnamefont {Wenner}}, \bibinfo {author} {\bibfnamefont {T.~C.}\
  \bibnamefont {White}}, \bibinfo {author} {\bibfnamefont {M.~R.}\ \bibnamefont
  {Geller}}, \bibinfo {author} {\bibfnamefont {A.~N.}\ \bibnamefont {Cleland}},
  \ and\ \bibinfo {author} {\bibfnamefont {J.~M.}\ \bibnamefont {Martinis}},\
  }\href@noop {} {\bibfield  {journal} {\bibinfo  {journal} {Phys. Rev. Lett.}\
  }\textbf {\bibinfo {volume} {113}},\ \bibinfo {pages} {220502} (\bibinfo
  {year} {2014})}\BibitemShut {NoStop}%
\bibitem [{\citenamefont {Wang}\ \emph {et~al.}(2001)\citenamefont {Wang},
  \citenamefont {S{\o}rensen},\ and\ \citenamefont {M{\o}lmer}}]{WangPRL01}%
  \BibitemOpen
  \bibfield  {author} {\bibinfo {author} {\bibfnamefont {X.}~\bibnamefont
  {Wang}}, \bibinfo {author} {\bibfnamefont {A.}~\bibnamefont {S{\o}rensen}}, \
  and\ \bibinfo {author} {\bibfnamefont {K.}~\bibnamefont {M{\o}lmer}},\
  }\href@noop {} {\bibfield  {journal} {\bibinfo  {journal} {Phys. Rev. Lett.}\
  }\textbf {\bibinfo {volume} {86}},\ \bibinfo {pages} {3907} (\bibinfo {year}
  {2001})}\BibitemShut {NoStop}%
\bibitem [{\citenamefont {Yang}\ and\ \citenamefont {Han}(2005)}]{YangPRA05b}%
  \BibitemOpen
  \bibfield  {author} {\bibinfo {author} {\bibfnamefont {C.-P.}\ \bibnamefont
  {Yang}}\ and\ \bibinfo {author} {\bibfnamefont {S.}~\bibnamefont {Han}},\
  }\href@noop {} {\bibfield  {journal} {\bibinfo  {journal} {Phys. Rev. A}\
  }\textbf {\bibinfo {volume} {72}},\ \bibinfo {pages} {032311} (\bibinfo
  {year} {2005})}\BibitemShut {NoStop}%
\bibitem [{\citenamefont {Lin}\ \emph {et~al.}(2006)\citenamefont {Lin},
  \citenamefont {Zhou}, \citenamefont {Ye}, \citenamefont {Xiao},\ and\
  \citenamefont {Guo}}]{LinPRA06}%
  \BibitemOpen
  \bibfield  {author} {\bibinfo {author} {\bibfnamefont {X.-M.}\ \bibnamefont
  {Lin}}, \bibinfo {author} {\bibfnamefont {Z.-W.}\ \bibnamefont {Zhou}},
  \bibinfo {author} {\bibfnamefont {M.-Y.}\ \bibnamefont {Ye}}, \bibinfo
  {author} {\bibfnamefont {Y.-F.}\ \bibnamefont {Xiao}}, \ and\ \bibinfo
  {author} {\bibfnamefont {G.-C.}\ \bibnamefont {Guo}},\ }\href@noop {}
  {\bibfield  {journal} {\bibinfo  {journal} {Phys. Rev. A}\ }\textbf {\bibinfo
  {volume} {73}},\ \bibinfo {pages} {012323} (\bibinfo {year}
  {2006})}\BibitemShut {NoStop}%
\bibitem [{\citenamefont {Yang}\ \emph {et~al.}(2010)\citenamefont {Yang},
  \citenamefont {Liu},\ and\ \citenamefont {Nori}}]{YangPRA10}%
  \BibitemOpen
  \bibfield  {author} {\bibinfo {author} {\bibfnamefont {C.-P.}\ \bibnamefont
  {Yang}}, \bibinfo {author} {\bibfnamefont {Y.-X.}\ \bibnamefont {Liu}}, \
  and\ \bibinfo {author} {\bibfnamefont {F.}~\bibnamefont {Nori}},\ }\href@noop
  {} {\bibfield  {journal} {\bibinfo  {journal} {Phys. Rev. A}\ }\textbf
  {\bibinfo {volume} {81}},\ \bibinfo {pages} {062323} (\bibinfo {year}
  {2010})}\BibitemShut {NoStop}%
\bibitem [{\citenamefont {Waseem}\ \emph {et~al.}(2012)\citenamefont {Waseem},
  \citenamefont {Irfan},\ and\ \citenamefont {Qamar}}]{WaseemPhysicaC12}%
  \BibitemOpen
  \bibfield  {author} {\bibinfo {author} {\bibfnamefont {M.}~\bibnamefont
  {Waseem}}, \bibinfo {author} {\bibfnamefont {M.}~\bibnamefont {Irfan}}, \
  and\ \bibinfo {author} {\bibfnamefont {S.}~\bibnamefont {Qamar}},\
  }\href@noop {} {\bibfield  {journal} {\bibinfo  {journal} {Physica C}\
  }\textbf {\bibinfo {volume} {477}},\ \bibinfo {pages} {24} (\bibinfo {year}
  {2012})}\BibitemShut {NoStop}%
\bibitem [{\citenamefont {Song}\ \emph {et~al.}(2012)\citenamefont {Song},
  \citenamefont {Shi}, \citenamefont {Xiang},\ and\ \citenamefont
  {Chen}}]{SongPhysicaB12}%
  \BibitemOpen
  \bibfield  {author} {\bibinfo {author} {\bibfnamefont {K.-H.}\ \bibnamefont
  {Song}}, \bibinfo {author} {\bibfnamefont {Z.-G.}\ \bibnamefont {Shi}},
  \bibinfo {author} {\bibfnamefont {S.-H.}\ \bibnamefont {Xiang}}, \ and\
  \bibinfo {author} {\bibfnamefont {X.-W.}\ \bibnamefont {Chen}},\ }\href@noop
  {} {\bibfield  {journal} {\bibinfo  {journal} {Physica B}\ }\textbf {\bibinfo
  {volume} {407}},\ \bibinfo {pages} {3596} (\bibinfo {year}
  {2012})}\BibitemShut {NoStop}%
\bibitem [{\citenamefont {Yang}\ \emph {et~al.}(2014)\citenamefont {Yang},
  \citenamefont {Su}, \citenamefont {Zhang},\ and\ \citenamefont
  {Shi-Biao~Zheng}}]{YangOL14}%
  \BibitemOpen
  \bibfield  {author} {\bibinfo {author} {\bibfnamefont {C.-P.}\ \bibnamefont
  {Yang}}, \bibinfo {author} {\bibfnamefont {Q.-P.}\ \bibnamefont {Su}},
  \bibinfo {author} {\bibfnamefont {F.-Y.}\ \bibnamefont {Zhang}}, \ and\
  \bibinfo {author} {\bibfnamefont {S.-B.}\ \bibnamefont {Shi-Biao~Zheng}},\
  }\href@noop {} {\bibfield  {journal} {\bibinfo  {journal} {Optics Letters}\
  }\textbf {\bibinfo {volume} {39}},\ \bibinfo {pages} {3312} (\bibinfo {year}
  {2014})}\BibitemShut {NoStop}%
\bibitem [{\citenamefont {Liu}\ \emph {et~al.}(2014)\citenamefont {Liu},
  \citenamefont {Fang}, \citenamefont {Liao},\ and\ \citenamefont
  {Liu}}]{LiuPRA14}%
  \BibitemOpen
  \bibfield  {author} {\bibinfo {author} {\bibfnamefont {X.}~\bibnamefont
  {Liu}}, \bibinfo {author} {\bibfnamefont {G.-Y.}\ \bibnamefont {Fang}},
  \bibinfo {author} {\bibfnamefont {Q.-H.}\ \bibnamefont {Liao}}, \ and\
  \bibinfo {author} {\bibfnamefont {S.-T.}\ \bibnamefont {Liu}},\ }\href@noop
  {} {\bibfield  {journal} {\bibinfo  {journal} {Phys. Rev. A}\ }\textbf
  {\bibinfo {volume} {90}},\ \bibinfo {pages} {062330} (\bibinfo {year}
  {2014})}\BibitemShut {NoStop}%
\bibitem [{\citenamefont {Harrow}\ \emph {et~al.}(2009)\citenamefont {Harrow},
  \citenamefont {Hassidim},\ and\ \citenamefont {Lloyd}}]{HarrowPRL09}%
  \BibitemOpen
  \bibfield  {author} {\bibinfo {author} {\bibfnamefont {A.~W.}\ \bibnamefont
  {Harrow}}, \bibinfo {author} {\bibfnamefont {A.}~\bibnamefont {Hassidim}}, \
  and\ \bibinfo {author} {\bibfnamefont {S.}~\bibnamefont {Lloyd}},\
  }\href@noop {} {\bibfield  {journal} {\bibinfo  {journal} {Phys. Rev. Lett.}\
  }\textbf {\bibinfo {volume} {103}},\ \bibinfo {pages} {150502} (\bibinfo
  {year} {2009})}\BibitemShut {NoStop}%
\bibitem [{\citenamefont {Clader}\ \emph {et~al.}(2013)\citenamefont {Clader},
  \citenamefont {Jacobs},\ and\ \citenamefont {Sprouse}}]{CladerPRL13}%
  \BibitemOpen
  \bibfield  {author} {\bibinfo {author} {\bibfnamefont {B.~D.}\ \bibnamefont
  {Clader}}, \bibinfo {author} {\bibfnamefont {B.~C.}\ \bibnamefont {Jacobs}},
  \ and\ \bibinfo {author} {\bibfnamefont {C.~R.}\ \bibnamefont {Sprouse}},\
  }\href@noop {} {\bibfield  {journal} {\bibinfo  {journal} {Phys. Rev. Lett.}\
  }\textbf {\bibinfo {volume} {110}},\ \bibinfo {pages} {250504} (\bibinfo
  {year} {2013})}\BibitemShut {NoStop}%
\bibitem [{\citenamefont {M\"ott\"onen}\ \emph {et~al.}(2004)\citenamefont
  {M\"ott\"onen}, \citenamefont {Vartiainen}, \citenamefont {Bergholm},\ and\
  \citenamefont {Salomaa}}]{MottonenPRL04}%
  \BibitemOpen
  \bibfield  {author} {\bibinfo {author} {\bibfnamefont {M.}~\bibnamefont
  {M\"ott\"onen}}, \bibinfo {author} {\bibfnamefont {J.~J.}\ \bibnamefont
  {Vartiainen}}, \bibinfo {author} {\bibfnamefont {V.}~\bibnamefont
  {Bergholm}}, \ and\ \bibinfo {author} {\bibfnamefont {M.~M.}\ \bibnamefont
  {Salomaa}},\ }\href@noop {} {\bibfield  {journal} {\bibinfo  {journal} {Phys.
  Rev. Lett.}\ }\textbf {\bibinfo {volume} {93}},\ \bibinfo {pages} {130502}
  (\bibinfo {year} {2004})}\BibitemShut {NoStop}%
\bibitem [{\citenamefont {Cai}\ \emph {et~al.}(2013)\citenamefont {Cai},
  \citenamefont {Weedbrook}, \citenamefont {Su}, \citenamefont {Chen},
  \citenamefont {Gu}, \citenamefont {Zhu}, \citenamefont {Li}, \citenamefont
  {Liu}, \citenamefont {Lu},\ and\ \citenamefont {Pan}}]{CaiPRL13}%
  \BibitemOpen
  \bibfield  {author} {\bibinfo {author} {\bibfnamefont {X.-D.}\ \bibnamefont
  {Cai}}, \bibinfo {author} {\bibfnamefont {C.}~\bibnamefont {Weedbrook}},
  \bibinfo {author} {\bibfnamefont {Z.-E.}\ \bibnamefont {Su}}, \bibinfo
  {author} {\bibfnamefont {M.-C.}\ \bibnamefont {Chen}}, \bibinfo {author}
  {\bibfnamefont {M.}~\bibnamefont {Gu}}, \bibinfo {author} {\bibfnamefont
  {M.-J.}\ \bibnamefont {Zhu}}, \bibinfo {author} {\bibfnamefont
  {L.}~\bibnamefont {Li}}, \bibinfo {author} {\bibfnamefont {N.-L.}\
  \bibnamefont {Liu}}, \bibinfo {author} {\bibfnamefont {C.-Y.}\ \bibnamefont
  {Lu}}, \ and\ \bibinfo {author} {\bibfnamefont {J.-W.}\ \bibnamefont {Pan}},\
  }\href@noop {} {\bibfield  {journal} {\bibinfo  {journal} {Phys. Rev. Lett.}\
  }\textbf {\bibinfo {volume} {110}},\ \bibinfo {pages} {230501} (\bibinfo
  {year} {2013})}\BibitemShut {NoStop}%
\bibitem [{\citenamefont {Pan}\ \emph {et~al.}(2014)\citenamefont {Pan},
  \citenamefont {Cao}, \citenamefont {Yao}, \citenamefont {Li}, \citenamefont
  {Ju}, \citenamefont {Chen}, \citenamefont {Peng}, \citenamefont {Kais},\ and\
  \citenamefont {Du}}]{PanPRA14}%
  \BibitemOpen
  \bibfield  {author} {\bibinfo {author} {\bibfnamefont {J.}~\bibnamefont
  {Pan}}, \bibinfo {author} {\bibfnamefont {Y.}~\bibnamefont {Cao}}, \bibinfo
  {author} {\bibfnamefont {X.}~\bibnamefont {Yao}}, \bibinfo {author}
  {\bibfnamefont {Z.}~\bibnamefont {Li}}, \bibinfo {author} {\bibfnamefont
  {C.}~\bibnamefont {Ju}}, \bibinfo {author} {\bibfnamefont {H.}~\bibnamefont
  {Chen}}, \bibinfo {author} {\bibfnamefont {X.}~\bibnamefont {Peng}}, \bibinfo
  {author} {\bibfnamefont {S.}~\bibnamefont {Kais}}, \ and\ \bibinfo {author}
  {\bibfnamefont {J.}~\bibnamefont {Du}},\ }\href@noop {} {\bibfield  {journal}
  {\bibinfo  {journal} {Phys. Rev. A}\ }\textbf {\bibinfo {volume} {89}},\
  \bibinfo {pages} {022313} (\bibinfo {year} {2014})}\BibitemShut {NoStop}%
\bibitem [{\citenamefont {Barz}\ \emph {et~al.}(2014)\citenamefont {Barz},
  \citenamefont {Kassal}, \citenamefont {Ringbauer}, \citenamefont {Lipp},
  \citenamefont {Daki\'c}, \citenamefont {Aspuru-Guzik},\ and\ \citenamefont
  {Walther}}]{BarzSciRep14}%
  \BibitemOpen
  \bibfield  {author} {\bibinfo {author} {\bibfnamefont {S.}~\bibnamefont
  {Barz}}, \bibinfo {author} {\bibfnamefont {I.}~\bibnamefont {Kassal}},
  \bibinfo {author} {\bibfnamefont {M.}~\bibnamefont {Ringbauer}}, \bibinfo
  {author} {\bibfnamefont {Y.~O.}\ \bibnamefont {Lipp}}, \bibinfo {author}
  {\bibfnamefont {B.}~\bibnamefont {Daki\'c}}, \bibinfo {author} {\bibfnamefont
  {A.}~\bibnamefont {Aspuru-Guzik}}, \ and\ \bibinfo {author} {\bibfnamefont
  {P.}~\bibnamefont {Walther}},\ }\href@noop {} {\bibfield  {journal} {\bibinfo
   {journal} {Sci. Rep.}\ }\textbf {\bibinfo {volume} {4}},\ \bibinfo {pages}
  {6115} (\bibinfo {year} {2014})}\BibitemShut {NoStop}%
\bibitem [{\citenamefont {Zheng}\ \emph {et~al.}(2017)\citenamefont {Zheng},
  \citenamefont {Song}, \citenamefont {Chen}, \citenamefont {Xia},
  \citenamefont {Liu}, \citenamefont {Guo}, \citenamefont {Zhang},
  \citenamefont {Xu}, \citenamefont {Deng}, \citenamefont {Huang},
  \citenamefont {Wu}, \citenamefont {Yan}, \citenamefont {Zheng}, \citenamefont
  {Lu}, \citenamefont {Pan}, \citenamefont {Wang}, \citenamefont {Lu},\ and\
  \citenamefont {Zhu}}]{ZhenPRL17}%
  \BibitemOpen
  \bibfield  {author} {\bibinfo {author} {\bibfnamefont {Y.}~\bibnamefont
  {Zheng}}, \bibinfo {author} {\bibfnamefont {C.}~\bibnamefont {Song}},
  \bibinfo {author} {\bibfnamefont {M.-C.}\ \bibnamefont {Chen}}, \bibinfo
  {author} {\bibfnamefont {B.}~\bibnamefont {Xia}}, \bibinfo {author}
  {\bibfnamefont {W.}~\bibnamefont {Liu}}, \bibinfo {author} {\bibfnamefont
  {Q.}~\bibnamefont {Guo}}, \bibinfo {author} {\bibfnamefont {L.}~\bibnamefont
  {Zhang}}, \bibinfo {author} {\bibfnamefont {D.}~\bibnamefont {Xu}}, \bibinfo
  {author} {\bibfnamefont {H.}~\bibnamefont {Deng}}, \bibinfo {author}
  {\bibfnamefont {K.}~\bibnamefont {Huang}}, \bibinfo {author} {\bibfnamefont
  {Y.}~\bibnamefont {Wu}}, \bibinfo {author} {\bibfnamefont {Z.}~\bibnamefont
  {Yan}}, \bibinfo {author} {\bibfnamefont {D.}~\bibnamefont {Zheng}}, \bibinfo
  {author} {\bibfnamefont {L.}~\bibnamefont {Lu}}, \bibinfo {author}
  {\bibfnamefont {J.-W.}\ \bibnamefont {Pan}}, \bibinfo {author} {\bibfnamefont
  {H.}~\bibnamefont {Wang}}, \bibinfo {author} {\bibfnamefont {C.-Y.}\
  \bibnamefont {Lu}}, \ and\ \bibinfo {author} {\bibfnamefont {X.}~\bibnamefont
  {Zhu}},\ }\href@noop {} {\bibfield  {journal} {\bibinfo  {journal} {Phys.
  Rev. Lett.}\ }\textbf {\bibinfo {volume} {118}},\ \bibinfo {pages} {210504}
  (\bibinfo {year} {2017})}\BibitemShut {NoStop}%
\bibitem [{\citenamefont {Lee}\ \emph {et~al.}(2019)\citenamefont {Lee},
  \citenamefont {Joo},\ and\ \citenamefont {Lee}}]{LeeSciRep19}%
  \BibitemOpen
  \bibfield  {author} {\bibinfo {author} {\bibfnamefont {Y.}~\bibnamefont
  {Lee}}, \bibinfo {author} {\bibfnamefont {J.}~\bibnamefont {Joo}}, \ and\
  \bibinfo {author} {\bibfnamefont {S.}~\bibnamefont {Lee}},\ }\href@noop {}
  {\bibfield  {journal} {\bibinfo  {journal} {Sci. Rep.}\ }\textbf {\bibinfo
  {volume} {9}},\ \bibinfo {pages} {4778} (\bibinfo {year} {2019})}\BibitemShut
  {NoStop}%
\end{thebibliography}%

\end{document}